\documentclass[11pt]{article}
\usepackage{moriond}
\usepackage[utf8]{inputenc}
\usepackage{amssymb} 
\usepackage{bbold}

\def\Journal#1#2#3#4{{#1} {\bf #2}, #3 (#4)}

\def\PLB{{\em Phys. Lett.}  B}
\def\PRL{\em Phys. Rev. Lett.}
\def\PRD{{\em Phys. Rev.} D}

\def\be{\begin{equation}}
\def\ee{\end{equation}}
\def\bea{\begin{eqnarray}}
\def\eea{\end{eqnarray}}

\newcommand{\Photo}{}

\begin{document}
\begin{flushright}
 LPT-Orsay-13-105
\end{flushright}

\vspace*{4cm}
\title{LEPTON UNIVERSALITY IN KAON DECAYS}

\author{ C. WEILAND }

\address{Laboratoire de Physique Théorique d'Orsay, Bâtiment 210, Université Paris Sud 11,\\ 91405 Orsay Cedex, France}

\maketitle
\abstracts{In the Standard Model extended by sterile neutrinos, modified $W\ell\nu$ couplings arise, which are able to induce a tree-level enhancement to lepton flavour universality violation in kaon decays. The additional mixing between the active neutrinos and the sterile ones can generate deviations from unitarity in the leptonic mixing matrix for charged currents. We reconsidered this idea in the context of the inverse seesaw and evaluated its impact on the well measured ratio $R_K$. We show that the current experimental bound can be saturated in agreement with the different experimental and observational constraints. Similar results can be obtained when considering the ratio $R_\pi$}

\section{Introduction}

Lepton flavour universality (LFU) is one of the distinctive features of the Standard Model (SM). Therefore, any deviation from the expected SM theoretical estimates in electroweak precision tests will signal the presence of New Physics. Here we focus on kaon leptonic decays which, in view of the expected experimental precision, have a unique potential to probe deviations from the SM regarding lepton universality.

In the SM, the dominant contribution to $\Gamma (K \rightarrow \ell \nu)$ arises from exchanges mediated by a $W$ boson. The prediction of each specific decay
is heavily plagued by hadronic matrix element uncertainties. However, in the ratio 
\begin{equation}\label{RK:Rpi}
R_K = \frac{\Gamma (K^+ \rightarrow e^+ \nu)}{\Gamma (K^+ \rightarrow \mu^+\nu)}\,,
\end{equation}
the hadronic uncertainties cancel out to a good approximation, so that the SM prediction can be computed with a high precision. In order to parametrize the deviation from the SM, it proves convenient to introduce a quantity, $\Delta r_K$, defined as 
\begin{equation}\label{deltar:P}
R_K = R_K^\mathrm{SM} (1+\Delta r_K)\,.
\end{equation}
The comparison of theoretical analyses~\cite{Cirigliano:2007xi,Finkemeier:1995gi} with the recent measurements from the NA62 collaboration~\cite{Lazzeroni:2012cx}
\begin{equation}\label{RK:Rpi:SMvsexp}
R_K^\mathrm{SM} = (2.477 \pm 0.001) \times 10^{-5}\,, \quad \quad R_K^\mathrm{exp} = (2.488 \pm 0.010) \times 10^{-5}\,,
\end{equation}
suggests that observation agrees at the $1 \sigma$ level with the SM's predictions, constraining the deviation to be rather small
\begin{equation}\label{deltar:P:value} 
\Delta r_K  =  (4 \pm 4 ) \times 10^{-3}\,. 
\end{equation}
The current experimental uncertainty on $\Delta r_K$ (of around 0.4\%) should be further reduced in the near future, as one expects to have $\delta R_K / R_K \sim0.1\%$~\cite{Goudzovski:2012gh}, which translates into measuring deviations $\Delta r_K \sim\mathcal{O}(10^{-3})$.

\section{$\Delta r_K$ in the presence of sterile neutrinos}

Once neutrino oscillations are incorporated into the SM, tree-level corrections to charged current interactions arise. In this case, charged weak interactions become non-diagonal:
\begin{equation}\label{cc-lag}
- \mathcal{L}_{cc} = \frac{g}{\sqrt{2}} U^{ji} \bar{l}_j \gamma^\mu P_L \nu_i  W_\mu^- + \, \mathrm{c.c.}\,,
\end{equation}
where $U$ is the product of the hermitian conjugate of the matrix diagonalizing the charged lepton mass matrix by the one that diagonalizes the neutrino mass matrix, $i = 1, \dots, n_\nu$ denoting the physical neutrino states and $j = 1, \dots, 3$ the charged leptons. In the case of three neutrino generations, $U$ corresponds to the unitary PMNS matrix. Because final state neutrinos cannot be experimentally distinguished, one has to sum over all $i=1,\dots,N_\mathrm{max}^{(\ell)}$, where $N_\mathrm{max}^{(l)}$ denotes the heaviest neutrino mass eigenstate which is kinematically allowed. Since we consider a tree-level process, long-distance corrections are neglected and the expression for $R_K$ is given by
\begin{equation}\label{RPresult}
R_K = \frac{\sum_i F^{i1} G^{i1}}{\sum_k F^{k2} G^{k2}}\,, \quad \mathrm{with}
\end{equation}
\begin{eqnarray}
 &F^{ij} = |U_{ji}|^2 \quad \mathrm{and}\nonumber \\  
 &G^{ij} = \left[m_K^2 (m_{\nu_i}^2+m_{l_j}^2) - (m_{\nu_i}^2-m_{l_j}^2)^2 \right] \left[ (m_K^2 - m_{l_j}^2 -
  m_{\nu_i}^2)^2 - 4 m_{l_j}^2 m_{\nu_i}^2 \right]^{1/2}\,.
  \label{FG}
\end{eqnarray}
A more detailed calculation can be found in~\cite{Abada:2012mc}. The result of Eq.~(\ref{RPresult}) has a straightforward interpretation: $F^{ij}$ represents the impact of the leptonic mixing (absent in the SM), whereas $G^{ij}$ encodes the mass-dependent factors.

Depending on the masses of the new states and most importantly, on their mixings with the active neutrinos, $\Delta r_K$ can considerably deviate from zero. In order to illustrate this, we consider two regimes: in the first (A), all sterile neutrinos are \textit{lighter} than the decaying kaon, but heavier than the active neutrino states, i.e. $m_\nu^\mathrm{active} \ll m_{\nu_{s}} \lesssim m_K$; in the second (B), all $\nu_{s}$ are \textit{heavier} than $m_K$. Notice that in case (A), all the mass eigenstates are kinematically accessible and one should sum over all $n_\nu$ states. Therefore, the enhancement to $\Delta r_K$ arises from different kinematic factors, see Eq.~(\ref{FG}). In scenario (B), the non-unitarity of the $3\times 3$ submatrix $\tilde U_\mathrm{PMNS}$ that relates the 3 charged leptons to the 3 active neutrinos leads to a violation of LFU. The deviations from unitarity of $\tilde U_\mathrm{PMNS}$ can be parametrized by the matrix $\eta$
\begin{equation}\label{U:eta:PMNS}
\tilde U_\mathrm{PMNS} = (\mathbb{1} - \eta)U_\mathrm{PMNS}\,.
\end{equation}
The impact of the non-unitarity of the lepton mixing matrix on leptonic light meson decays was first  investigated in~\cite{Shrock:1980vy,Shrock:1980ct}, prior to the confirmation of neutrino oscillations. In our work, we revisit this idea in the light of recent neutrino data and in view of the present (and future) experimental sensitivities to $\Delta r_{K}$~\cite{Lazzeroni:2012cx,Goudzovski:2012gh}.

\section{$\Delta r_K$ in the  inverse seesaw model}

As an illustrative example, we chose the inverse seesaw (ISS) to study the new contributions to $\Delta r_K$. In the ISS, the SM particle content is extended by $n_R$ generations of right-handed (RH) neutrinos  $\nu_R$ and $n_X$ generations of singlet fermions $X$ with equal lepton number. Here we will consider the case $n_R = n_X = 3$, but it worth noting that deviations from unitarity can occur for different values of $n_R$ and $n_X$. The Lagrangian is given by
\begin{equation}
 \mathcal{L}_\mathrm{ISS} = - \sum_{i,j} \left( Y^{ij}_\nu \overline{L_i} \widetilde{H} \nu_{Rj} + M_R^{ij} \overline{\nu_{Ri}} X_j + \frac{1}{2} \mu_{X}^{ij} \overline{X_{i}^C} X_{j} + h.c. \right)\,,
\label{L_IS}
\end{equation}
where $i,j = 1,2,3$ are generation indices, $\widetilde{H} = i \sigma_2 H^*$, $M_R$ and $\mu_X$ are mass matrices appearing in terms conserving and violating lepton number conservation, respectively. The distinctive feature of the ISS is that the additional $\mu_X$ parameter allows to accommodate the smallness of the active neutrino masses $m_\nu$ for a low seesaw scale, still having natural Yukawa couplings ($Y_\nu\sim {\mathcal{O}}(1) $). As a consequence, the mixing between active neutrinos and the additional sterile states, approximately given by $Y_\nu v / M_R$, can be sizeable.

We numerically evaluate the contributions to $R_K$ in the framework  of the ISS and address the two scenarios discussed before, which can be translated in terms of ranges for the (random) entries of $M_R$: {\it scenario (A)}, ${M_R}_{i} \in [0.1,200]$ MeV; {\it scenario (B)}, ${M_R}_{i} \in [1,10^6]$ GeV. The entries of $\mu_X$ have also been randomly varied in the $[0.01$ eV$, 1$ MeV$]$ range for both cases.

The adapted Casas-Ibarra parametrization for $Y_\nu$, which can be found in~\cite{Abada:2012mc}, ensures that neutrino oscillation data is satisfied (we use the best-fit values from a global analysis~\cite{Tortola:2012te}, and set the CP violating phases of $U_\mathrm{PMNS}$ to zero). We have also imposed the various constraints on sterile neutrinos as well as non-unitarity constraints that are relevant to our scenarios. A detailed discussion of those can be found in~\cite{Abada:2012mc}. In our numerical analysis, the $R$ matrix  angles are taken to be real (thus no contributions to lepton electric dipole moments are expected), and randomly varied in the range ${\theta}_{i} \in [0,2 \pi]$. Although we do not discuss it here, we have verified that similar $\Delta r_K$ contributions are found when considering the more general complex $R$ matrix case.

\begin{figure}[ht]
\begin{tabular}{cc}
\hspace*{11mm}{\footnotesize Scenario (A)} &
\hspace*{11mm}{\footnotesize Scenario (B)}\vspace*{2mm} \\
\includegraphics[width=0.47\textwidth]{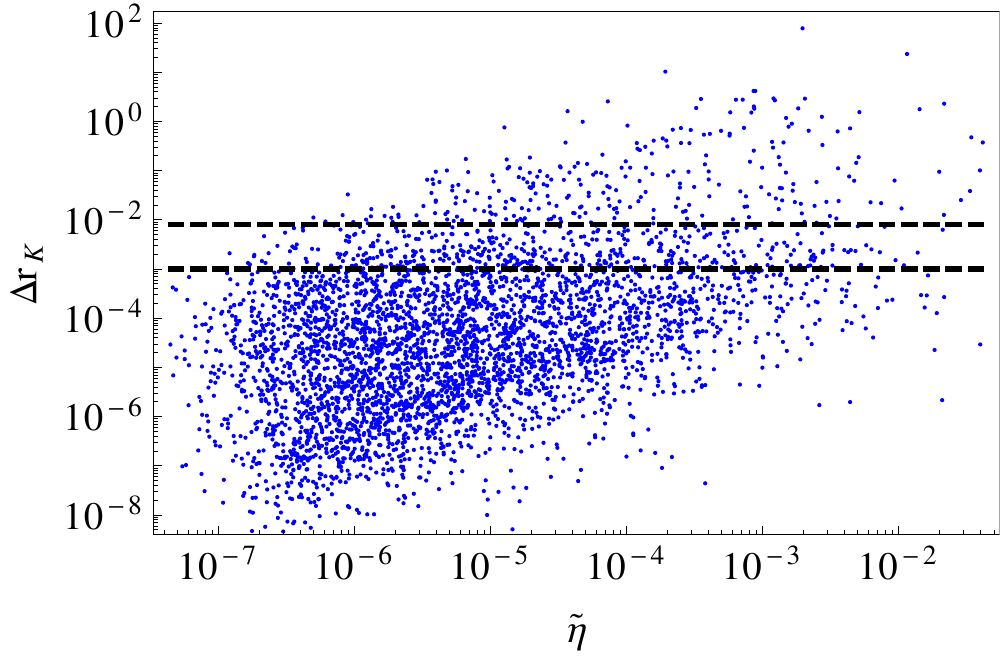}
&
\includegraphics[width=0.47\textwidth]{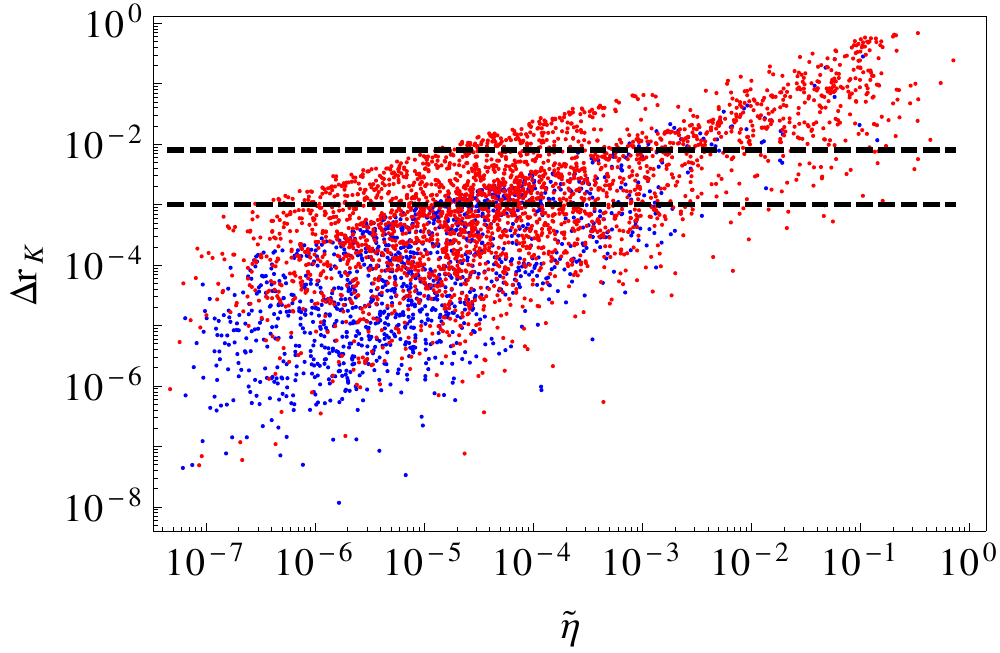} 
\end{tabular}
\caption{Contributions to $\Delta r_K$ in the inverse seesaw as a
  function of $\tilde \eta = 1 - |\mathrm{Det}(\tilde U_\mathrm{PMNS})|$:
  scenarios A (left) and B (right). The upper (lower) dashed line
  denotes the current experimental limit (expected
  sensitivity).  On the right panel, red points denote
    cases where $Y_\nu \ge 10^{-2}$.
 All points  comply with experimental and laboratory constraints. 
 Points in (B) are also in agreement with cosmological bounds, while those in (A) 
 require considering a non-standard cosmology.} 
\label{figure1}
\end{figure}

In Fig.~\ref{figure1}, we collect our results for $\Delta r_K$ in scenarios (A) - left panel - and (B) - right panel, as a function of $\tilde \eta$, which parametrizes the departure from unitarity of the active neutrino mixing submatrix $\tilde U_\mathrm{PMNS}$, $\tilde \eta = 1 - |\mathrm{Det}(\tilde U_\mathrm{PMNS})|$.
In scenario (A),
one can have very large contributions to $R_K$, which can even reach values $\Delta r_K \sim \mathcal{O}(1)$ or higher.  The hierarchy of the sterile neutrino spectrum in case (A) is such that one can indeed have a significant amount of LFU violation, while still avoiding non-unitarity bounds.  Although this scenario would in principle allow to produce sterile neutrinos in light meson decays, the smallness of the associated $Y_\nu$ ($\le\mathcal{O}(10^{-4})$), together with the loop function suppression, precludes this possibility and the observation of LFV processes.  The strong constraints from CMB and X-rays would exclude scenario (A); in order to render it viable, one would require a non-standard cosmology~\cite{Gelmini:2008fq}.

In case (B), sizeable LFU violation is also possible, with deviations from the SM predictions again as large as $\Delta r_K \sim \mathcal{O}(1)$. The large deviations in this scenario typically occur when all  the singlet states are considerably heavier than the decaying meson, and reflect specific features of the ISS: large active-sterile mixings can occur, thus leading to an enhancement of $R_K$.  Besides, the large $Y_\nu$ open the possibility of having larger contributions to LFV observables so that, for example, BR($\mu \to e \gamma)$ can be within MEG reach in this case.\\

To conclude this discussion, the existence of sterile neutrinos can potentially lead to a significant violation of lepton flavour universality at tree-level in kaon decays via a modified $W \ell \nu$ vertex. As an illustrative example, we have evaluated the contributions to $R_K$ in the SM extended by the inverse seesaw mechanism. Our analysis reveals that very large deviations from the SM predictions can be found ($\Delta r_K \sim \mathcal{O}(1))$, or even larger, well above the limit of the NA62 experiment at CERN. This makes $R_K$ a potentially constraining observable for models with sterile neutrinos.  We further notice that these large deviations are a generic and non fine-tuned feature of this model. Finally, it should be added that the same results can be obtained when considering pion decays and the ratio $R_\pi$~\cite{Abada:2012mc}.

\section*{Acknowledgements}

The author acknowledges financial support from the Moriond organizing committee to attend the conference. This work has been partly done under the ANR project CPV-LFV-LHC NT09-508531.

\end{document}